\documentclass[%
 reprint,
 amsmath,amssymb,
 aps,
]{revtex4-2}
\usepackage{comment}
\usepackage{graphicx}% Include figure files
\usepackage{dcolumn}% Align table columns on decimal point
\usepackage{bm}% bold math
\begin{document}

\preprint{APS/123-QED}

\title{Observation of elastic bound modes in the continuum in architected beams}% Force line breaks with \\

\author{Adib Rahman}
\author{Raj Kumar Pal}%
 \email{rkpal@ksu.edu}
\affiliation{%
 Department of Mechanical and Nuclear Engineering, Kansas State University, Manhattan, Kansas 66506, USA\\
 }%

\date{\today}% It is always \today, today,

\begin{abstract}
We report the experimental observation of an elastic bound mode in the continuum (BIC) in a compact region of an architected beam. We consider a long slender beam with rigid masses attached at periodic intervals, with a compact segment bounded by four protruding side beams. The key idea is to seek a mode where the side beams move out-of-phase with the compact region, thereby nullifying the forces and moments outside this region and resulting in a bound mode. The structure is modeled using Euler-Bernoulli beam theory and the side beams are designed by imposing equilibrium constraints required for a BIC. Multiple BICs are found in the compact region, and for each BIC, we find a one-parameter family of BIC supporting side beam designs. The predictions are verified by three-dimensional finite element simulations, followed by their experimental observation using laser Doppler vibrometry in a macro-scale structure. Our approach allows to achieve BICs in an arbitrary sized compact region of the architected beam. Our findings may open avenues for confining elastic wave energy in compact regions for applications in sensors and resonators.
\end{abstract}

\maketitle

%\tableofcontents

\section{Introduction}
Bound modes in the continuum (BICs) are a unique class of localized modes with two key properties: their wave amplitude diminishes to zero outside a compact region and their frequency is in the continuous spectrum (pass band) of bulk propagating modes~\cite{hsu2016bound}. In contrast, conventional bound modes reside within bandgaps, and the localized modes encountered in the passband typically exhibit leakage, with the wave amplitude gradually decreasing from the center of the wave~\cite{jo2020graded,khelif2003trapping,marchal2012dynamics}. The concept of BICs originated in quantum mechanics, introduced by von Neumann and Wigner in 1929, who utilized a complex artificial potential~\cite{neumann1993merkwurdige}. It was regarded as a mathematical anomaly, as such complex potentials were not possible in real materials. BICs were subsequently predicted and observed in several classical wave systems~\cite{FONDA1963123,PhysRevA.11.446}. Notably, in 1966, BICs were experimentally observed in acoustics through the `wake shedding experiment'~\cite{parker1966resonance}. Today, BICs have become an active area of research across various scientific disciplines due to their leak free energy storing capacity with very high quality factor (Q factor)~\cite{KOSHELEV2019836}. Potential applications of BICs encompass  lasing~\cite{imada1999coherent,hirose2014watt,kodigala2017lasing}, sensing~\cite{Romano:19}, filtering~\cite{Doskolovich:19,foley2014symmetry}, supersonic surface acoustic device~\cite{kawachi2001optimal,naumenko2003surface}, vibration absorption~\cite{cao2021perfect}, and wave guiding~\cite{benabid2002stimulated,couny2007generation,rahman2022bound}.

Recent advancements in manufacturing techniques have opened up new possibilities for exploring BICs in complex structures, particularly in the domains of photonic metamaterials. Two major types of BICs are symmetry-protected and accidental. Symmetry-protected BICs arise from the mismatch between the spatial symmetry of a localized mode and the symmetry of the propagating modes. Experimental observations of such symmetry-protected BICs have been reported in various systems, such as dielectric slabs with square arrays of cylinders~\cite{hsu2013observation}, periodic chains of dielectric disks~\cite{sadrieva2019experimental}, and optical waveguides~\cite{plotnik2011experimental}. On the other hand, accidental BICs can be achieved through precise system parameter tuning to cancel their coupling with bulk  propagating waves. One example of this category is the Fabry-Perot BIC~\cite{PhysRevLett.100.183902}, where the BIC is formed through the destructive interference of waves. In addition to these two types of BICs, recent research has explored quasi-BICs (QBICs) which have high Q factors~\cite{taghizadeh2017quasi,abujetas2021high}. As true BIC-supporting structures are limited, quasi-BICs are emerging as an alternative.

In contrast to photonics, a major challenge in achieving elastic BICs is the simultaneous presence of transverse and longitudinal waves with distinct dispersion relations. BICs should not couple or hybridize with any propagating modes present in an elastic body. There have been a few works in recent years on predicting and observing BICs in elastic media. Examples include prediction of elastic BICs in a structure comprising of two periodic arrays of cylinders at a specific distance~\cite{haq2021bound}, observing BICs in a chain of thin plates connected by slender beams by exploiting non-Hermitian effects~\cite{fan2022observation}. BICs have also been observed in multi-physics domains, including in chip scale ring-shaped optomechanical microresonators~\cite{yu2022observation}, slab-on-substrate phononic crystals~\cite{tong2020observation}, elastic bar with air-encapsulated cavity~~\cite{lee2023elastic}. Cao  \textit{et al.}\cite{cao2021perfect,cao2021elastic} observed quasi-BICs in a semi-infinite plate attached to a resonant waveguide and predicted these can be turned to BICs by tuning the geometric parameters~\cite{cao2021elastic}. All these BICs require specific material properties, boundary conditions, geometric features and dimensions. For practical applications, it is desirable to have a general framework that can translate across material properties and generate BICs in arbitrary sized compact regions.

This work builds on our prior work~\cite{rahman2022bound}, where we predicted how a family of BICs can be achieved in an arbitrary compact region of a spring-mass system by exploiting symmetry constraints. Here, we extend this concept to realize BICs in a compact region of an architected beam. In contrast to spring-mass chains, beams are continuous structures with multiple degrees of freedom at each point, namely transverse displacements and rotations. These degrees of freedom impose additional conditions for BICs. Here, we consider a periodic architected beam having an array of rigid masses. To achieve BICs in a compact region, four side beams are added, and the key idea is that they move out-of-phase with the periodic beam to nullify forces and moments at their joints. 

The outline of the paper is as follows: section~\ref{sec:designModel} presents the design and modeling approach. The structure is modeled using $1D$ Euler-Bernoulli beam theory. Section~\ref{sec:numer_soln} presents the BIC mode shapes determined using $1D$ finite element analysis and reports a 1-parameter family of BIC supporting side beam designs. In section~\ref{sec:expts}, $3D$ finite element simulations and laser Doppler vibrometry based experimental measurements are presented, that verify and validate the existence of a BIC in the architected structure. The simulations are done using the beam theory-based design as a starting point to finalize a structure that simplifies fabrication. Finally, the conclusions, along with various sources of error and possible future extensions are presented in section~\ref{sec:conc}.

\section{Proposed concept and modeling approach to design compact region}\label{sec:designModel}

We first introduce the proposed architected beam and discuss the key idea of achieving BICs in a compact region by adding side beams. These side beams are designed by modeling the structure using a one-dimensional ($1D$) Euler Bernoulli beam theory. A description of this modeling approach is presented, followed by its numerical discretization procedure.

\subsection{Architected beams with side segments and symmetry consideration} \label{sec:design}

\begin{figure*}
    \centering
    \includegraphics[width = 12cm]{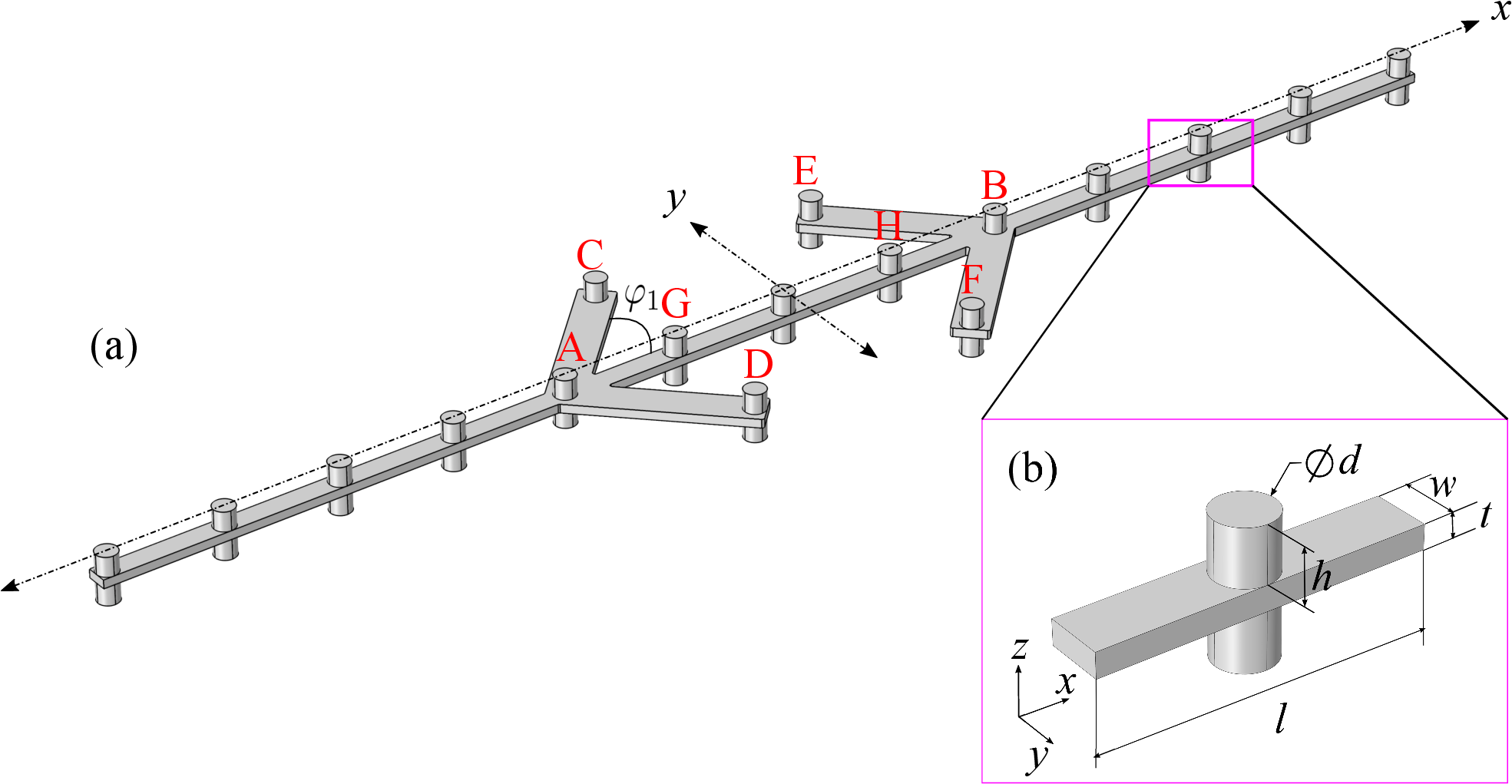}
    \caption{(a) Schematic of proposed architected beam: rigid masses attached at periodic intervals along a homogeneous beam. Four side beams are added to get a BIC between A and B. 
    (b) A unit cell of the periodic beam with the key geometric variables labeled.}
    \label{fig:model}
\end{figure*}

Let us consider a homogeneous beam with rigid masses attached at periodic intervals of distance $l$. An example is shown in the central beam in Fig.~\ref{fig:model}(a). We call this periodic architected  beam as the main beam. Figure~\ref{fig:model}(b) displays a  unit cell of the main beam with the key geometric variables labeled. It is a slender beam with rectangular cross-section and has two identical rigid cylinders at its center, one each at the top and bottom. 

Our objective is to achieve a BIC in an arbitrary compact region, for example between the cross-sections labeled A and B in Fig.~\ref{fig:model}(a). We will model this architected structure using one-dimensional ($1D$) beam theory, which assumes that the beam deforms such that each cross-section remains rigid. Under this assumption, the degrees of freedom are the 3 translations and 3 rotations of each cross-section along the beam's axis. We restrict attention to long wavelengths, compared to the beam thickness and low frequencies, i.e., in the first pass band of the beam. The lower frequency band has flexural modes with displacement along $z$. For such modes, it suffices to consider two degrees of freedom at each cross section: transverse displacement $u$ along $z$ and rotation $\theta$ about y-axis, with the latter accounting for bending. The presence of side beams couples the torsional (rotation about $x$) and flexural modes near the compact region. However, as we discuss below, for BIC modes, the rotation about $x$ is cancelled due to symmetry and thus, the number of relevant DOFs at each point along the beam cross-section is two.
 
A BIC between sections $A$ and $B$ will be a mode with displacement confined in this region and with zero force and moment at sections $A$ and $B$. Note that if the net force and net moment on section $A$ ($B$) is zero, it will be at rest and no displacement field will be induced to the left (right) of point $A$ ($B$). Our approach is to add side beams so that the sections $A$ and $B$ are at rest and we will have BIC in the compact region. Let us discuss how the four side beams in Fig.~\ref{fig:model}(a) induce BICs and the reason for having two side beams on either side the main beam at sections $A$ and $B$.  The key idea is to have a mode where the side beams move out of phase, i.e., in opposite direction to the main beam, thereby cancelling the net force and moment at sections $A$ and $B$. A single side beam will induce torsional rotation about the main beam axis due to the component of moment along $x$. To cancel this moment, a second side beam is added. The two side beams on either side at $A$ thus move in-phase with each other, but out-of-phase with the main beam.  We will show  later in Sec.~\ref{sec:numer_soln} a family of side beams that can achieve exact cancellation of forces and moments. 

We choose all the side beams to be identical and arrange them so that the center of the compact region between sections $A$ and $B$ has reflection symmetry about both $x$ and $y$ axes. Note that BICs do not require these symmetries and they are chosen to simplify the side beam design. Indeed, the only requirement is that the force $F_z=0$ and moments $M_x = M_y = 0$ at sections $A$ and $B$. This requirement ensures zero displacement and rotation at these two sections, and consequently, outside the compact region. We remark here that the full design space of distinct side beams is sufficiently large, with multi-parameter families of solutions that satisfy these conditions. Imposing the constraints arising for symmetry, the problem of inducing BICs reduces to determining suitable side beams. As we are considering that the side beams' arrangement is reflection symmetric about the $x$ and $y$ axis, determining one side beam's design will suffice to complete the beam structure that can support BIC at a particular frequency. 

Let us discuss how these reflection symmetries and equilibrium conditions impose restrictions on resulting bound mode shapes in the compact region. Each symmetry can be represented by a linear transformation operator. This operator maps the position vectors of each point in the structure to its corresponding reflected point. In addition, the mode shapes are eigenvectors of this operator~\cite{dresselhaus2007group}. The reflection symmetry operator has two eigenvalues: $\lambda = \pm 1$ and the bound mode shapes are thus even ($\lambda = 1$) or odd ($\lambda=-1$)  in the compact region about the symmetry axis. Let us  analyze the consequence of reflection symmetry about $x$ axis. An odd mode shape about the $x$ axis will induce a moment and thus rotation about $x$ at sections $A$ and $B$ as the side arms move in opposite directions. The sections to the left of $A$ and right of $B$ will thus not be at rest and a bound mode is thus not possible with an odd mode shape about the $x$ axis. In summary, a bound mode shapes in the compact region will be either even or odd about the $y$ axis and even about the $x$ axis.  

\subsection{Modeling with Euler-Bernoulli beam theory and numerical procedure} \label{sec:1dBeamModel}
Let us derive the governing equations for free vibrations of the structure based on $1D$ beam theory and discuss the finite element based procedure to solve them. Let $u(x,t) $ and $u_p(x,t)$ denote the transverse displacements of the main beam and side beam $p$, respectively. The action functional for this structure is given by 
\begin{widetext}
\begin{multline}\label{euler_bernoulli_beam} 
S = \int_{0}^T \int_{0}^{L} \left[
\dfrac{\rho A \dot{u}^2 }{2} 
- 
\dfrac{EI \left(u^{\prime\prime}\right)^2}{2} 
+ 
\sum_{p = 1}^N \left( \dfrac{m \dot{u}^2}{2}  + \dfrac{I_r \dot{\theta}^2 }{2}\right) \delta(x - p l) \right] dx dt  \\
+
\sum_{p=1}^4 \dfrac{1}{\cos \varphi_p}\int_{0}^T \int_{0}^{L_p} \left[ 
\dfrac{\rho A \dot{u}_p^2}{2}
- 
\dfrac{\cos^4 \varphi_p  EI \left(u_p^{\prime\prime}\right)^2}{2} 
+ 
\left( \dfrac{m_p \dot{u}_p^2}{2} 
+ 
\dfrac{\cos^2 \varphi_p I_{rp} \dot{\theta}_p^2}{2}\right) \delta(x - L_p) \right]dx dt
\end{multline}
\end{widetext}
Here $u^\prime$ and $\dot{u}$ denote partial derivatives of $u$ with respect to $x$ and $t$, respectively, $\theta=u^\prime$ is the rotation of the section and $L_p/\cos\varphi_p$ is the length of side beam $p$. $E$, $I$, $\rho$ and $A$ are the Young's modulus,  bending moment of inertia, density and cross-section area, respectively. The attached cylinders are assumed to be rigid with diameter $d$. The bending moment of inertia is  $I =  wt^3/12$ for a beam with width $w$ and thickness $t$. $\varphi_p$ is angle of side beam $p$ with respect to the $x$-axis. We seek harmonic solutions at frequency $\omega$ and impose a displacement field of the form $u(x,t) = u(x)e^{i\omega t}$ to replace the time derivatives by $i \omega$. The displacement field satisfies the Euler Lagrange equations, obtained by setting variation of $S$ to zero. This condition gives 
\begin{widetext}
\begin{multline}
    \delta S = \int_0^L \left[-\omega^2 
    \left( \rho A u  \delta u + \sum_{p=1}^N  (m u \delta u + I_r u^\prime \delta u^\prime ) \delta(x-pl) \right)- EI u^{\prime\prime} \delta u^{\prime\prime}\right]  \; dx   \\ 
+  
\sum_{p=1}^4 \dfrac{1}{\cos \varphi_p}\int_0^{L_p} \left[ -\omega^2 \left( \rho A u_p \delta u_p + \left( m_p u_p \delta u_p + \cos^2 \varphi_p I_{rp} u^\prime_p \delta u^\prime_p \right) \delta(x- L_p) \right) -\cos^4 \varphi_p EI u_p^{\prime\prime} \delta u_p^{\prime\prime}\right] \; dx = 0   . \label{eq:d_action}
\end{multline}
\end{widetext}
Now, let us discuss the numerical procedure to discretize and solve the above equation. We use a finite element approximation, where unknown degrees of freedom are restricted to be the displacements $u$ and rotations $u^\prime$ at the locations of attached masses. We express $u$ and $u^\prime$ at a point in the structure as a weighted sum of piecewise cubic  polynomials, i.e., having continuous first derivatives, and the weights being the degrees of freedom. We seek a solution that satisfies the governing equation~\eqref{eq:d_action} for any perturbation fields $\delta u$ and $\delta u^\prime$ that lies in the same space spanned by the degrees of freedom. Explicit expressions for the polynomials and the resulting equations are presented in the appendix. The resulting discretized eigenvalue problem for the structure may be written in the matrix form as 
\begin{equation}
    \omega^2\bm{Mu} = \bm{Ku}\label{eqn:gov_mat} . 
\end{equation}
Here $\bm{u}$ is the vector of unknown degrees of freedom, i.e., displacements and rotations at masses, and $\bm{M}$, $\bm{K}$ are the discretized mass and stiffness matrices, respectively. 

\begin{figure}[!t]
    \centering
    \includegraphics[width = 6.5cm]{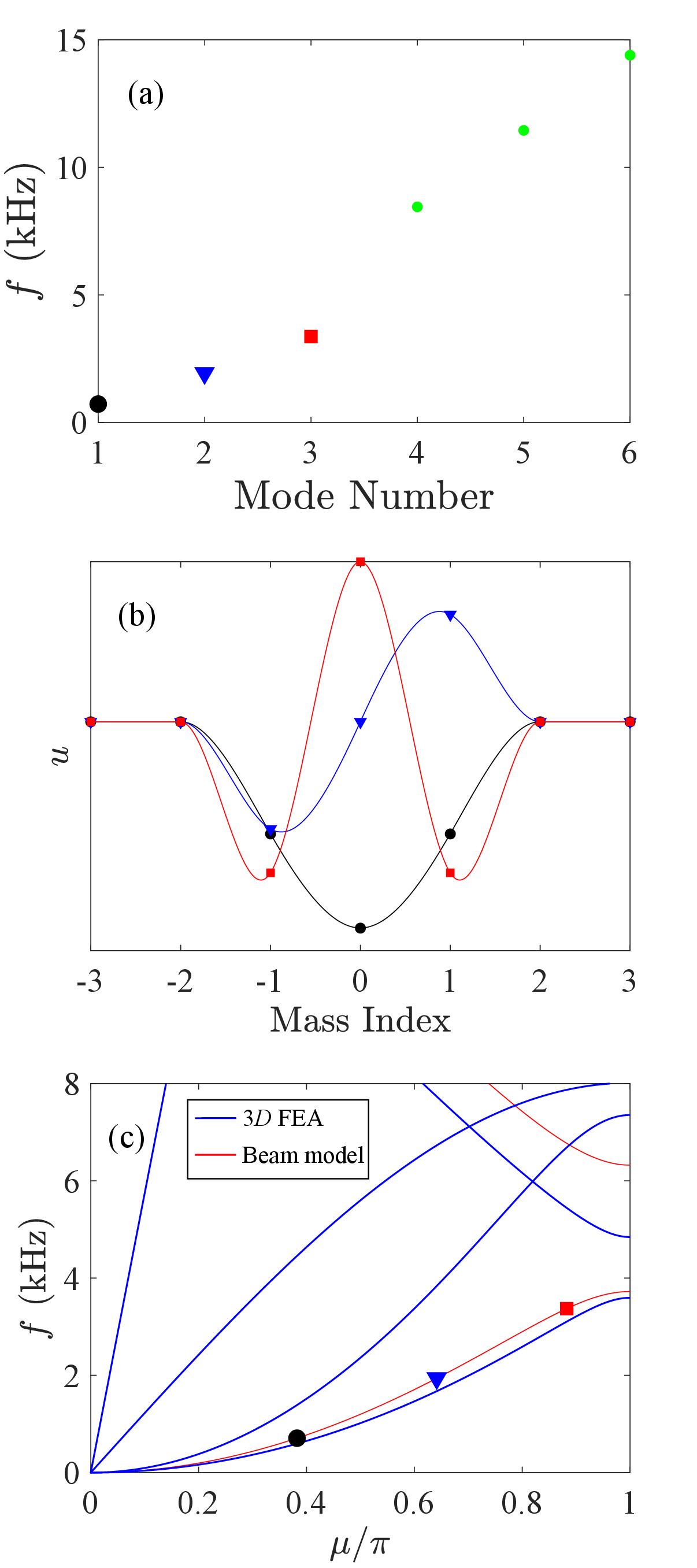}
    \caption{(a) Frequencies of bound modes in the compact region of Fig.~\ref{fig:model}(a). (b) Mode shapes of first 3 modes, $u$ is  transverse displacement along the main beam. Markers indicate rigid mass locations, with similar marker in (a) for their frequency. (c) Dispersion diagram of the main beam from $1D$ beam theory and $3D$ elasticity. The first three mode frequencies lie on the lowest flexural band and are thus BICs. }
    \label{fig:BIC}
\end{figure}

\section{Numerical solution of architected beams supporting BICs}\label{sec:numer_soln}

In this section, the mode shapes of BICs and a family of side beams that support these BICs are determined for a given main beam. Although our studies are presented for a specific choice of compact region, the concept and approach can be extended to an arbitrary sized compact region and material properties. A two step process is used to design BIC supporting structures using the beam model introduced above in Sec.~\ref{sec:1dBeamModel}. The first step is to determine the bound mode frequencies by imposing zero displacement and rotation at sections $A$ and $B$ in the main beam.   The next step is to determine the side beam dimensions that satisfy the equilibrium conditions required to keep sections $A$ and $B$ at rest. Finally, we verify if these modes are indeed BICs, i.e., if their frequency lie in a pass band. This is done by performing a dispersion analysis that yields the pass and stop band frequencies. 

\subsection{Bound mode frequencies and design of side beams}\label{sec:BIC_freq}

Having derived the governing equations for the proposed structure, let us now solve them numerically to determine BICs. The first step is to determine the bound modes by considering the compact region of the main beam only and explicitly enforcing zero displacement and rotation at sections $A$ and $B$ in Fig.~\ref{fig:model}(a). The resulting modes will, in general, not satisfy equilibrium conditions at sections $A$ and $B$. The second step is to determine  the side beam dimensions so that the total structure (main and side beams together) satisfy the equilibrium conditions at sections $A$ and $B$. In these mode shapes, sections $A$ and $B$ will thus be at rest and have zero net force and moment, thereby ensuring that parts outside the compact region will be at rest. Thus, this procedure ensures bound modes in the structure.  

In the first step, to determine the natural frequency and mode shape of BICs, we use material properties of Aluminum 6061 with Young's modulus $(E = 68.9\;\text{GPa}$, Poisson's ratio $\nu = 0.3$ and density $\rho = 2700\;\text{kg/m}^3)$ for the beam and Neodymium Magnet N35 $( \text{cylinder density}, \rho_c = 7537.6\;\text{kg/m}^3)$ for the cylinders considering ease of fabrication. The key geometric variables $(l,w,t,d,h)$ in the unit cell, see Fig.~\ref{fig:model}(b) are chosen to be (27.5 mm, 5 mm, 2.032mm, 5mm, 4.6mm). The bound mode shapes and  frequencies are determined by solving the eigenvalue problem~\eqref{eqn:gov_mat} in the compact region with zero displacement and rotation boundary conditions. Figure~\ref{fig:BIC}(a) displays the six bound mode frequencies. 

Next, we need to determine the side beam dimensions so that the structure supports BIC in the compact region. There are several geometric variables for the side beams as shown in Fig.~\ref{fig:model}(b) for a unit cell as well as the angle between the main the beam and side beam $AC$, $\varphi_1$ as shown in Fig.~\ref{fig:model}(a). Different sets of the geometric variables for side beams can give BICs in the compact region. We fix $\varphi$ as 45$^\circ$ to simplify the problem. Also, for ease of fabrication, the beam thickness and the cylinder diameters in the side beams are chosen to be the same as that in the main beam. Thus the design reduces to determining three geometric variables: length ($l_s$), width ($w_s$) of the side beams, and the cylinder height ($h_C$) at section $C$ in Fig.~\ref{fig:model}(a). 

Let us summarize the conditions on a side beam displacement field $u_p$ needed to get a BIC. These conditions ensure that section $A$ and the region to the left of it will be at rest. Recall the key idea that the two identical side beams at section $A$, as in Fig.~\ref{fig:model}(a) move out of phase with the main beam, thereby canceling the force and moment at $A$. Its displacement field has to satisfy the governing equations~\eqref{eq:d_action} at the bound mode frequency $\omega$ under fixed boundary conditions at $A$ ($u_p = 0$ and $u^\prime_p=0$). In addition, the resulting forces and moments from the side and main beams should add to zero so that section $A$ is in equilibrium. Under the considered $1D$ beam theory, the force and moment at section $A$ are given by 
\begin{align*}
    F &= EI u^{\prime\prime \prime} +  \sum_{p=1}^2\cos^3 \varphi_p\;u_p^{\prime \prime \prime} ,\\
    M &= EI u^{\prime\prime} + \sum_{p=1}^2\cos^2 \varphi_p\; E I_{p} u_p^{\prime \prime } . 
\end{align*}
Now, we derive the discrete approximations of the above  conditions for the side beam having section $C$ in Fig.~\ref{fig:model}(a). 
A side beam is modeled using a single finite element and the degrees of freedom are the displacements and rotations at the two ends ($A$ and  $C$). Since we seek solutions with section $A$ fixed, the displacement field simplifies to  $u_p(x)=N_3(x/l_s) \theta_C + N_4(x/l_s) u_C$. Explicit expressions for $N_3$, $N_4$ are presented in appendix.  Under this approximation, the governing equations of side beams  and the equilibrium conditions at $A$ then reduce to 
%\begin{widetext}
\begin{subequations}\label{eqn:sidebeams}
\begin{multline}
    \delta \theta_C = 0 
    \implies \left( \dfrac{4EI_s}{l_s} - \dfrac{\omega^2m_s}{420}(4l_s^2+I_C)\right) \theta_C + \\
    \left( \dfrac{11\omega^2m_s}{210}l_s - \dfrac{6EI_s}{l_s^2}\right) u_C = 0, \label{eqn:NR1}
\end{multline}
\begin{multline}
        \delta u_C = 0 
    \implies \left( \dfrac{11\omega^2m_s}{210}l_s - \dfrac{6EI_s}{l_s^2}\right) \theta_C + \\
    \left( \dfrac{12EI}{l_s^3} - \dfrac{\omega^2m_s}{420}(156+m_C)\right) u_C = 0,\label{eqn:NR2}
\end{multline}
    \begin{multline}
     F_A = 0 
    \implies \dfrac{4EI_s}{l_s}\theta_C - \dfrac{12EI_s}{l_s^2}u_C  \\ - \dfrac{2EI}{l_e}\theta_G + \dfrac{6EI}{l_e^2}u_G = 0, \label{eqn:NR3}    
    \end{multline}
    \begin{multline}
     M_A = 0 
    \implies \dfrac{12EI_s}{l_s^2}\theta_C - \dfrac{24EI_s}{l_s^3}u_C - \\
    \dfrac{6EI}{l_e^2}\theta_G 
    + \dfrac{12EI}{l_e^3}u_G = 0 . \label{eqn:NR4}
    \end{multline}
\begin{comment}
\begin{align}
    &\delta \theta_C = 0\nonumber \\
    &\implies \left( \dfrac{4EI_s}{l_s} - \dfrac{\omega^2m_s}{420}(4l_s^2+I_C)\right) \theta_C + \nonumber\\
    &\left( \dfrac{11\omega^2m_s}{210}l_s - \dfrac{6EI_s}{l_s^2}\right) u_C = 0, %\label{eqn:NR1}
    \\    
    \nonumber\\ 
    &\delta u_C = 0 \nonumber\\ 
    &\implies \left( \dfrac{11\omega^2m_s}{210}l_s - \dfrac{6EI_s}{l_s^2}\right) \theta_C +\nonumber\\
    &\left( \dfrac{12EI}{l_s^3} - \dfrac{\omega^2m_s}{420}(156+m_C)\right) u_C = 0, %\label{eqn:NR2}
    \\
    \nonumber\\ 
    &F_A = 0 \nonumber\\
    &\implies \dfrac{4EI_s}{l_s}\theta_C - \dfrac{12EI_s}{l_s^2}u_C - \dfrac{2EI}{l_e}\theta_G + \dfrac{6EI}{l_e^2}u_G = 0,  %\label{eqn:NR3}
    \\
    \nonumber\\ 
    &M_A = 0 \nonumber\\
    &\implies \dfrac{12EI_s}{l_s^2}\theta_C - \dfrac{24EI_s}{l_s^3}u_C - \dfrac{6EI}{l_e^2}\theta_G + \dfrac{12EI}{l_e^3}u_G = 0 . %\label{eqn:NR4}
\end{align}
\end{comment}
\end{subequations}
%\end{widetext}
Here $m_s = \rho l_s w_s t$ and $I_s = w_s t^3/12$ are mass and bending moment of inertia of the side beam, respectively, while $m_C= \pi\rho_c d^2 h_C/4 $ and $I_C = m_C/12(3d^2/4+h_C^2) $ are mass and mass moment of inertia of the cylindrical mass at section $C$. $u_G$ and $\theta_G$ are the displacement and rotation at section $G$, corresponding to the mode shape of the compact region at frequency $\omega$ (see Fig.~\ref{fig:BIC}(a,b)). The force and moment balance constraints assume that the two side beams at $A$ move in phase. Indeed, as discussed earlier, a BIC mode shape is symmetric about the $x$-axis. 

\begin{figure}[!t]
    \centering
    \includegraphics[height = 6cm]{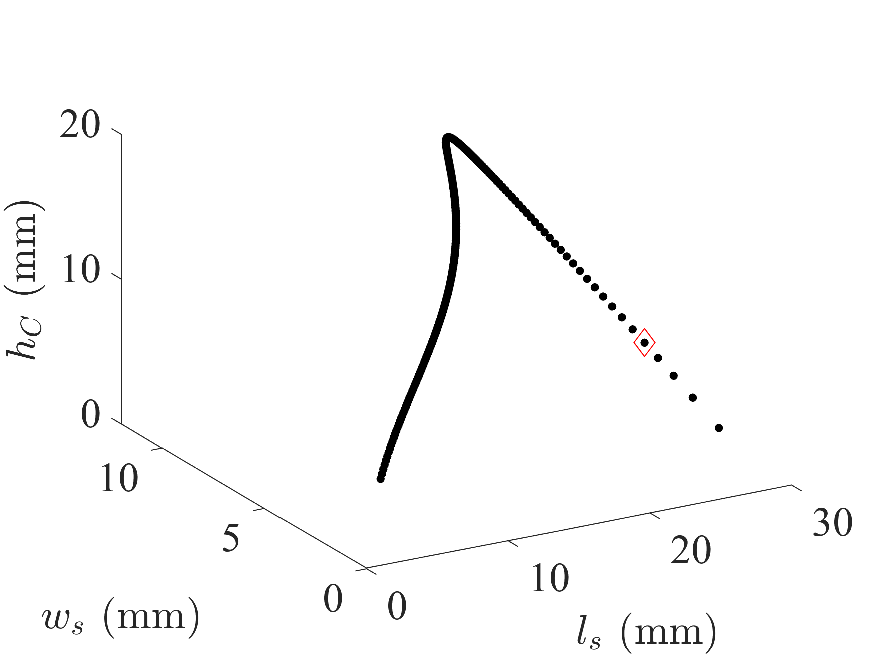}
    \caption{A 1-parameter family of side beams support the lowest frequency BIC (black marker in Fig.~\ref{fig:BIC}(b)). The diamond marked geometric dimensions are used for experimental demonstration in Sec.~\ref{sec:expts}.} 
    \label{fig:map}
\end{figure}

The conditions for getting a BIC mode lead to a system of four nonlinear equations~\eqref{eqn:sidebeams} with five unknown variables $(u_C,\theta_C,l_s, w_s,h_C)$ related to the side beams. To determine them, $l_s$ is set to different fixed values in a wide range and the remaining variables are determined using the Newton-Raphson method. We determined side beam dimensions that support the lowest frequency bound mode at 713 Hz, denoted by a black marker in Fig.~\ref{fig:BIC}(a). Figure~\ref{fig:map} displays a 1-parameter family of solutions that we obtained as $l_s$ is varied. Side beams for every solution in Fig.~\ref{fig:map} induce the bound mode shown by the black curve in Fig.~\ref{fig:BIC}(b). Similarly for the frequencies marked by blue triangle and red square, we find families of design parameters which support the corresponding bound modes in Fig.~\ref{fig:BIC}(b). These design parameter are displayed in the appendix, Fig.~\ref{fig:appendix_map}. 

\subsection{Dispersion analysis of the architected beams}
To confirm if the bound modes in Fig.~\ref{fig:BIC}(b) are indeed BICs, i.e., if their  frequency lies in the pass band, we do a dispersion analysis of the main beam, which is periodic with unit cell in Fig.~\ref{fig:model}(b). We work with the discrete approximation, where the degrees of freedom are $(u_n, \theta_n)$ at a section having rigid mass labeled $n$. We seek traveling wave solutions of the form $\bm{u}_n = \Tilde{\bm{u}}e^{i\kappa n}$, where  $\kappa$ is the non-dimensional wave-number, $\bm{u}_n = [\theta_n,\; u_n]^T$ and ${\Tilde{\bm{u}}} =[\theta\;u]^T$. The discretized governing equations~\eqref{eqn:gov_mat} for this section then reduce to an eigenvalue problem  $\omega^2\bm{M}_n(\kappa)\Tilde{\bm{u}}=\bm{K}_n(\kappa)\Tilde{\bm{u}}$ with
\begin{widetext}
\begin{subequations}
\begin{align}
\bm{M}_n(\kappa) &=\dfrac{m_{b}}{420}\begin{bmatrix}
			             2l_e^2(4-3\cos\kappa) + {I_y}_n      &26l_e\sin\kappa \\
			             -26il_e\sin\kappa    &312+108\cos\kappa + m_n\\                  
		\end{bmatrix},\;
		\label{eqn:M_n}	\\   
\bm{K}_n(\kappa) &= \dfrac{EI}{l_e^2}\begin{bmatrix}
			             4l_e(2+\cos\kappa)      & -12i\sin\kappa\\ 
			             12i \sin\kappa    &{24}/{l_e}(1-\cos\kappa) 
			\end{bmatrix} . \label{eqn:K_n}
\end{align}
\end{subequations}
\end{widetext}
Solving the eigenvalue problem for each $\kappa$ in the interval [0, $\pi$] gives two dispersion branches denoted by red curves in the Fig.~\ref{fig:BIC}(c). The first three bound mode frequencies in Fig.~\ref{fig:BIC}(a) lie on the lower red branch, implying that these are BICs.

\section{$3D$ numerical simulations  and experimental results}\label{sec:expts}
This section presents the verification of our predictions based on the $1D$ beam model using $3D$ elasticity theory. The simulations are performed using a commercial finite element analysis software COMSOL. Finally, we report on the experimental observation of a BIC under dynamic excitation with a shaker. 

\begin{figure*}[!t]
    \centering
    \includegraphics[height = 5cm]{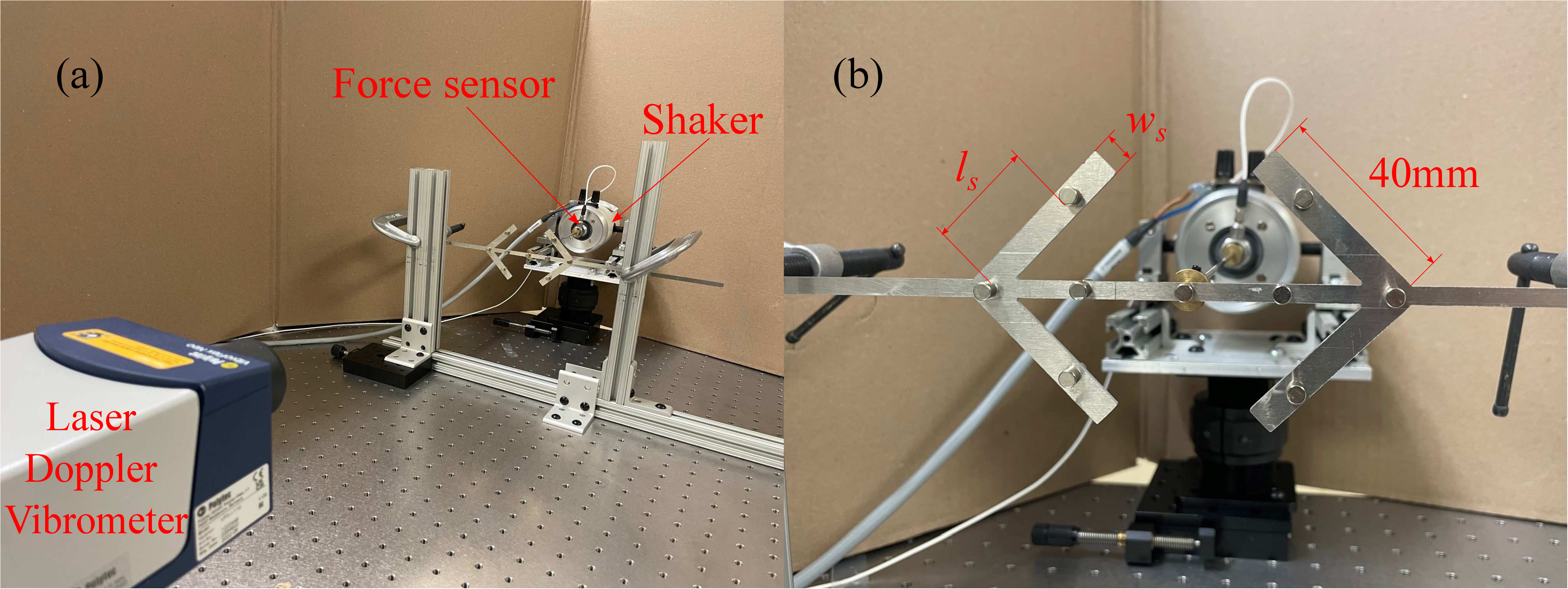}
    \caption{(a) Experimental set-up. The beam is exited at the center of the compact region using a shaker. A force sensor is attached to the shaker. The velocity at various points along the beam is measured using a laser vibrometer. (b) Zoomed-in view, indicating the dimensions $l_s$ and $w_s$ in a 40 mm side beam length.}
    \label{fig:exp}
\end{figure*}

\subsection{Verification using $3D$ elasticity theory}
We model the finite beam structure shown in Fig.~\ref{fig:exp}(b) using $3D$ elasticity theory. Here, the motion at every point in a structure made of a linear elastic isotropic solid is governed by~\cite{achenbach2012wave} $\rho \bm{\ddot{u}} - [(\lambda+\mu)\nabla(\nabla\cdotp\bm{u})+\mu\nabla^2\bm{u}] = 0$, with $\bm{u}=[u_x\;u_y\;u_z]^T$ being the vector of displacement components, $\mu$ and $\lambda$ the Lam\'{e} constants of the solid. A finite element analysis is performed using COMSOL Multiphysics software and the domain is discretized using quadratic elements with tetrahedral geometry. 

Let us first verify if the $1D$ beam model is accurate by comparing their corresponding dispersion surfaces for the unit cell of Fig.~\ref{fig:model}(b). Figure~\ref{fig:BIC}(c) displays this comparison, with the blue and red curves determined using the $3D$ and $1D$ models, respectively. The lower frequency flexural branch is quite close for the two models, which demonstrates the effectiveness of the $1D$ beam model in predicting flexural modes based BICs. In addition, the $3D$ analysis also shows a quadratic bending along the $y$-direction, linear longitudinal and torsional dispersion branches.

Let us now determine the final design using $3D$ finite element analysis. All cylinders attached to the main and side beams are taken to be identical to simplify fabrication. Our starting design point is indicated by the diamond marker in Fig.~\ref{fig:map}, with side beam dimensions $(l_s,w_s,h_C) =$ (28 mm, 5.81 mm, 5.37 mm). We do a detailed $3D$ analysis and make minor changes to the design predicted using the $1D$ beam theory. There are two reasons for requiring modifications to the design predicted using the $1D$ model. First, the rigid masses are assumed to be a point mass and the space occupied due to the finite diameter $d$ is neglected in the $1D$ model. The second reason is to simplify assembly of masses in the side beams, the side beam is made longer to $40$ mm and masses are attached at a distance $l_s$, as shown in Fig.~\ref{fig:exp}(b). This design is distinct from the $1D$ beam model, where the cylindrical mass on the side beams are attached at their ends. 

\begin{figure*}[!t]
    \centering
    \includegraphics[height = 10cm]{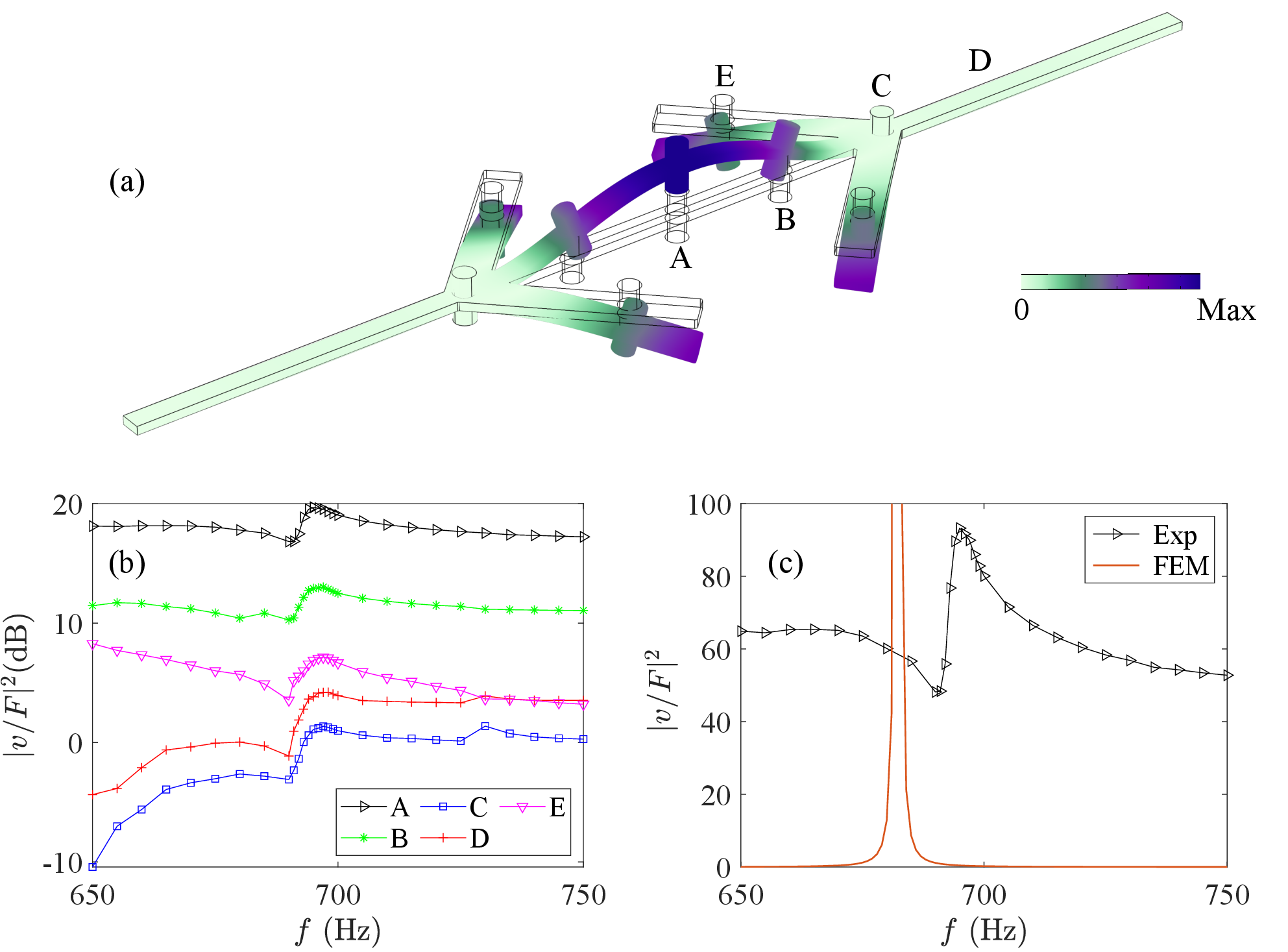}
    \caption{(a) BIC mode shape at 682.5 Hz determined using $3D$ FEA. (b) Measured frequency response at cylinders. A resonant peak occurs at 695 Hz corresponding to the BIC. (c) Comparison of frequency response at excitation point $A$: simulation and experiments.}
    \label{fig:comsol}
\end{figure*}

We search for suitable $l_s$ and $w_s$ by doing a parametric sweep over these variables near the starting design point using $3D$ finite element simulations.  This choice is guided by the existence of 1-parameter family of valid design solutions of Eqn.~\eqref{eqn:sidebeams}, see Fig.~\ref{fig:map}. The sweep search yields $(l_s,w_s) =$ (28 mm, 8.23 mm) when the height and diameter of all the attached cylinders are set to $h=4.6$ mm and $d=5$ mm, respectively. Figure~\ref{fig:comsol}(a) displays the BIC mode shape confined to the compact region for these side beam dimensions. Note that, the frequency determined with the $3D$ model (682.5 Hz)  is close to that predicted by the $1D$ model (713 Hz). 

\subsection{Experimental observation of a BIC}
Finally, we report on the experimental observation of the BIC shown in Fig.~\ref{fig:comsol}(a) using the dimensions determined from the $3D$ analysis. The experimental setup is shown in Fig.~\ref{fig:exp}(a). The structure in Fig.~\ref{fig:exp}(b) is fabricated from a 0.08 inch thick Aluminum 6061 sheet using water-jet cutting. Cylindrical Neodymium magnets N35 of 5mm diameter and 4.6mm height are placed at the top and bottom in the main and side beams. These dimensions are chosen since they are commercially available. The sample is clamped at both ends and a permanent magnet shaker (LDS V203) is used to apply sinusoidal displacement at the center of the compact region. The excitation point is denoted here by $A$, as shown in Fig.~\ref{fig:comsol}(a). A force sensor (PCB 208C01) is attached to the shaker to measure the applied force. The velocity at various points along the main beam is measured using a  laser Doppler vibrometer  (Polytec VFX-I-110).

Let us summarize the experimental procedure. We excite the structure at different frequencies in the interval $[650, 750]$ Hz and determine the frequency response function. The excitation frequencies are indicated by markers in the response plot in Fig.~\ref{fig:comsol}(a). The velocity of a cylinder and the force applied by the shaker are measured by applying the excitation at a frequency for 15 seconds. To allow transients to die down, the force and velocity data are recorded in the last 6 seconds of excitation. Then the beam is kept at rest for 10 seconds before exciting it at next frequency. This process is repeated for measuring velocity of each cylinder. The maximum velocity, $v$ and maximum force, $F$ at a frequency are calculated by a FFT of the measured velocity and force. Then, normalized energy of every point is determined as $|v/F|^2$.

Figure~\ref{fig:comsol}(b) displays the measured response at cylinders on the right side of excitation point. We observe a peak in the frequency response of the excitation point $A$ at 695 Hz. At this frequency, the response at points $C$ and $D$, lying at the boundary and outside the compact region, are significantly lower ($1$ unit) compared to the excitation point $A$ (about 93 units). This observation confirms the existence of a BIC in the structure at 695 Hz. We also measured the frequency response at other cylinders left of the excitation point. Note that due to reflection symmetry, the corresponding symmetric points on the left should have identical response. We quantified the difference in response at points $B$, $C$, $D$ and the side beams with their corresponding symmetric points to the left at the BIC frequency. Our experiments showed a 3$\%$ difference point B and at the side beams, and a 10$\%$ difference at points $C$ and $D$ with their corresponding symmetric points to the left. 

Figure~\ref{fig:comsol}(c) displays a comparison between experiments and simulations for the frequency response at the excitation point. The simulation has a response peak (around 1000 units) at 682.5 Hz, which is the frequency of the mode shape in Fig.~\ref{fig:comsol}(a). Let us remark on the various sources of error that result in deviation from a true BIC. The discrepancy between experiments and simulations is attributed to imperfections in manufacturing, with the fabricated beam width being around 4.8 mm instead of the designed 5mm, and precision in placing cylinders at the exact locations.  Another possible reason for lower peak response in experiments is air damping ~\cite{yu2022observation} at the macro scales. 

Finally, we note that the frequency response point $C$ is 0.137 units in simulations, compared to $1000$ for point $A$. Although significantly lower than the experimental value, it is not zero, in contrast to the prediction of the $1D$ beam model. To understand this discrepancy, note that the Euler Bernoulli beam theory, which is used to predict a true BIC, assumes that each cross-section is non-deformable, and undergoes translations and rigid rotations~\cite{gere1997mechanics}. Although an excellent approximation at low frequencies, small deviations arise when full $3D$ effects are considered. The resulting displacement at $C$ is thus a measure of the deviation of the exact $3D$ solution from the $1D$ beam theory. Notably, we do not attempt to satisfy the zero displacement condition point-wise, but instead do it in an average sense over the entire cross-section. 

\section{Conclusion}\label{sec:conc}

We introduce the architected structure that comprises of a main beam with periodically attached masses, and has a compact region with protruding side beams. Symmetry and equilibrium constraints are used to determine the conditions required for a BIC in this compact region. A $1D$ beam model is derived using Euler Bernoulli theory and a finite element method is used to determine bound modes in the structure. The conditions on main and side beams required to support a BIC are derived and a Newton Raphson method is used to solve the resulting nonlinear equations. For each BIC, we find a 1-parameter family of side beam designs that supports it. A dispersion analysis is conducted to confirm that their frequencies lie in passband and they are thus BICs. 

We verify the predictions of BIC based on the $1D$ beam model using finite element analysis (FEA) based on $3D$ elasticity. The $1D$ model is found to be in good agreement for the low frequency flexural modes under consideration. For ease of fabrication and assembly, and to account for the mismatch with $3D$ elasticity, minor modifications to the side beam design determined using the $1D$ model are made by doing a parametric sweep over the width and length of side beams using $3D$ FEA. The designed structure is fabricated and excited over a range of frequencies around the BIC frequency. 
The experimental results of frequency response at the excitation point shows a resonant peak close to the frequency predicted by FEA. At the resonant frequency, the fraction of energy leaking to surrounding reduced to minimum, which demonstrates the existence of a BIC in the compact region. The experimental results are compared with FEA results and the possible reasons for discrepancies, along with causes of deviation from a true BIC are discussed. 

Let us remark on some future possible extensions of our work. These concepts translate across length scales and material properties, and may find applications at the micro and nano-scales. At those scales, the limitations associated with air damping and material damping may be significantly reduced. The idea of cancelling forces and moments by exploiting symmetry may be extended to realize bound modes and BICs in plates, shells and $3D$ architected solids.

\begin{acknowledgments}
This work was supported by the U.S. National Science
Foundation under Award No. 2027455
\end{acknowledgments}

\appendix
\section{Finite element shape functions and matrices}

\begin{figure}[!bht]
    \centering
    \includegraphics[height = 4.5cm]{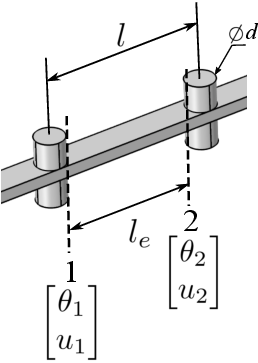}
    \caption{Schematic of a beam finite element with degrees of freedom  $[\theta_1, \;u_1]^T$ and $[\theta_2, \;u_2]^T$ at sections 1 and 2, respectively.}
    \label{fig:beam_elem}
\end{figure}

Let us discuss an approximation we use to determine the effective bending stiffness of the beam segment between sections $1$ and $2$ in Fig.~\ref{fig:beam_elem}. Its contribution comes from three segments: the two segments of length $d/2$ at the two ends having rigid masses, and the beam segment with length $l_e=l-d$ between them. These three segments may be viewed as springs in series and their effective stiffness is lower than the stiffness of these segments. The   segment having rigid masses (length $d/2$) has significantly higher bending stiffness and its contribution to the effective bending stiffness is neglected. Only the beam segment $l_e$ in Fig.~\ref{fig:beam_elem} is used to determine the effective bending stiffness. Under this approximation, the rigid masses may be represented as point masses with a beam segment of length $l_e$ between them. 

Figure~\ref{fig:beam_elem} displays a schematic of a beam finite element. The discrete degrees of freedom are the displacements and rotations at the location of rigid masses, labeled $[\theta_1,\;u_1]^T$ and $[\theta_2,\;u_2]^T$ in this figure. Let us denote the location of point 1 and 2 by $x_1$ and $x_2$, respectively. In this element, we seek a solution of the form 
\begin{equation}\label{eqn:u_fe}
   u(x) = N_1(\xi) \theta_1 + N_2(\xi) u_1 + N_3(\xi) \theta_2 + N_4(\xi) u_2, 
\end{equation}
where $\xi$ is a local coordinate in the element given by $(x-x_1)/l_e$ and taking values in $[0,1]$. $N_i(\xi)$ are Hermite polynomial shape functions~\cite{cook_concepts_2001} and explicit expressions for the shape functions are
\begin{subequations}\label{eqn:shape_functions}
\begin{align}
    N_1&=l_e\xi(\xi-1)^2\\
    N_2&=1-3\xi^2+2\xi^3\\
    N_3&=l_e\xi^2(\xi-1)\\
    N_4&=\xi^2(3-2\xi)
\end{align}
\end{subequations}
Here, Eqn.~\eqref{eqn:u_fe} may be written compactly as $u(x) = \bm{N} \bm{u}^T$, with $\bm{N}$ and $\bm{u}$ being vectors having components $N_i$ and $u_i$, respectively.  

Let us derive the contribution of the above beam segment to the governing equation. We substitute Eqn.~\eqref{eqn:u_fe} into Eqn.~\eqref{eq:d_action} and separate the terms with and without $\omega^2$ into mass matrix, $\bm{M}_{el}$ and stiffness matrix, $\bm{K}_{el}$ respectively for an element. The various terms in Eqn.~\eqref{eq:d_action} then have the form $\delta \bm{u}^T \bm{K}_{el} \bm{u}$ or $\omega^2  \delta \bm{u}^T \bm{M}_{el} \bm{u}$, where 
\begin{align}
 \bm{K}_{el} &= \int_{x_1}^{x_2}\dfrac{d^2\bm{N}^T}{dx^2}EI\dfrac{d^2\bm{N}}{dx^2}dx , \\
\bm{M}_{el} &= \int_{x_1}^{x_2} \rho A \bm{N}^T\bm{N}dx \nonumber\\
&+ \sum^2_{i = 1}m_i\bm{N}^T(\xi_i)\bm{N}(\xi_i) +\sum^2_{i = 1}I_i\dfrac{d\bm{N}^T(\xi_i)}{dx}\dfrac{d\bm{N}(\xi_i)}{dx} .
 \end{align}
    Here, $x_1$ and $x_2$ are respectively locations for section $1$ and $2$ in Fig.~\ref{fig:beam_elem}. The $m_i$ and $I_i$ represent mass and mass moment of inertia of $i^{th}$ rigid mass, respectively. 
Explicit expressions for these matrices are 
%\begin{widetext}

\begin{subequations}
\begin{align}
&\bm{M}_{el} =\dfrac{m_{b}}{420} \begin{bmatrix}
			 4l_e^2 + I_1 &22l_e   &-3l_e^2   &13l_e\\
			 22l_e    &156 + m_1     &-13l_e    &54\\
			 -3l_e^2  &-13l_e  &4l_e^2 + I_{2}    &-22l_e\\
			 13l_e   &54      &-22l_e    &156+m_{2}\\
		\end{bmatrix} 
\label{eqn:M}
\\
&\bm{K}_{el} = \dfrac{EI}{l_e^3} \begin{bmatrix}
			 4 l_e^2       & 6 l_e   & 2 l_e^2   & -6 l_e\\
			 6 l_e  & 12  & 6 l_e & -12 \\
              2 l_e^2     &6 l_e   & 4 l_e^2    &-6 l_e\\
              -6 l_e   & -12 & -6 l_e & 12 \\
			\end{bmatrix}\label{eqn:K}.
\end{align}
\end{subequations}
    
%\end{widetext}
Here, $m_b=\rho A l_e$ is the mass of the beam in the  element. Assembling $\bm{M}_{el}$ and $\bm{K}_{el}$ for every beam segment in the structure gives its  mass $\bm{M}$ and  stiffness $\bm{K}$ matrix. The governing equations result in an eigenvalue problem $\omega^2\bm{M} \bm{u} = \bm{K} \bm{u}$. Here, $\bm{u}$ is a vector with components having the displacements and rotations at the rigid mass locations.

\section{1-parameter family of designs for other BICs}

\begin{figure}[!h]
    \centering
    \includegraphics[height = 11cm]{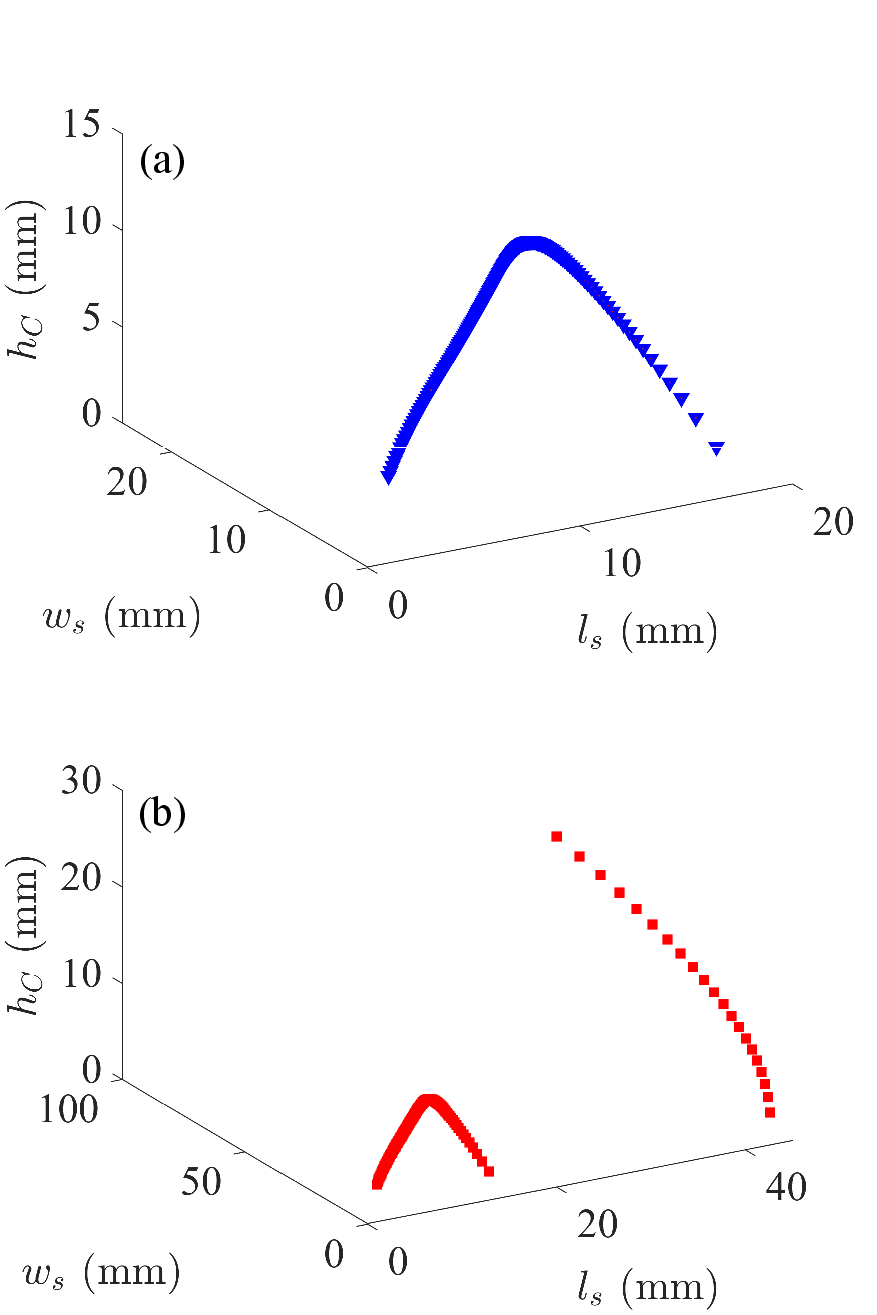}
    \caption{1-parameter family of side beam design that support the BICs marked by blue (a) and red (b) markers in Fig.~\ref{fig:BIC}(b), respectively. The curve in (b) is continuous. It looks discontinuous as the design variables take negative values (not shown). }
    \label{fig:appendix_map}
\end{figure}

%\nocite{*}

\bibliography{main}% Produces the bibliography via BibTeX.
\end{document}